\documentclass[twoside,a4paper,11pt]{sea22}
\usepackage{graphicx}
\usepackage{hyperref}
\usepackage{movie15}
\usepackage{color}
\usepackage{siunitx}  
\usepackage{natbib}    
\usepackage[export]{adjustbox}
\def\hoy{\number\day \space de \space\ifcase\month\or
 Enero\or Febrero\or Marzo\or Abril\or Mayo\or Junio\or
 Julio\or Agosto\or Septiembre\or Octubre\or Noviembre\or Diciembre\fi
 \space de \number\year}
\def\ii/{\'{\i}}
\def\cion/{ci\'on}
\def\cao/{\c c\~ao}
\def\utw{\smash{\rlap{\lower5pt\hbox{$\sim$}}}}
\def\udtw{\smash{\rlap{\lower6pt\hbox{$\approx$}}}}

\def\tens#1{\ifmmode\mathchoice{\mbox{$\sf\displaystyle#1$}}
{\mbox{$\sf\textstyle#1$}}
{\mbox{$\sf\scriptstyle#1$}}
{\mbox{$\sf\scriptscriptstyle#1$}}\else
\hbox{$\sf\textstyle#1$}\fi}
\def\vec#1{\ifmmode\mathchoice{\mbox{\boldmath$\displaystyle#1$}}
{\mbox{\boldmath$\textstyle#1$}}
{\mbox{\boldmath$\scriptstyle#1$}}
{\mbox{\boldmath$\scriptscriptstyle#1$}}\else
\hbox{\boldmath$\textstyle#1$}\fi}
\def\bbbc{{\mathchoice {\setbox0=\hbox{$\displaystyle\rm C$}\hbox{\hbox
to0pt{\kern0.4\wd0\vrule height0.9\ht0\hss}\box0}}
{\setbox0=\hbox{$\textstyle\rm C$}\hbox{\hbox
to0pt{\kern0.4\wd0\vrule height0.9\ht0\hss}\box0}}
{\setbox0=\hbox{$\scriptstyle\rm C$}\hbox{\hbox
to0pt{\kern0.4\wd0\vrule height0.9\ht0\hss}\box0}}
{\setbox0=\hbox{$\scriptscriptstyle\rm C$}\hbox{\hbox
to0pt{\kern0.4\wd0\vrule height0.9\ht0\hss}\box0}}}}
\def\bbbq{{\mathchoice {\setbox0=\hbox{$\displaystyle\rm
Q$}\hbox{\raise
0.15\ht0\hbox to0pt{\kern0.4\wd0\vrule height0.8\ht0\hss}\box0}}
{\setbox0=\hbox{$\textstyle\rm Q$}\hbox{\raise
0.15\ht0\hbox to0pt{\kern0.4\wd0\vrule height0.8\ht0\hss}\box0}}
{\setbox0=\hbox{$\scriptstyle\rm Q$}\hbox{\raise
0.15\ht0\hbox to0pt{\kern0.4\wd0\vrule height0.7\ht0\hss}\box0}}
{\setbox0=\hbox{$\scriptscriptstyle\rm Q$}\hbox{\raise
0.15\ht0\hbox to0pt{\kern0.4\wd0\vrule height0.7\ht0\hss}\box0}}}}
\def\bbbt{{\mathchoice {\setbox0=\hbox{$\displaystyle\rm
T$}\hbox{\hbox to0pt{\kern0.3\wd0\vrule height0.9\ht0\hss}\box0}}
{\setbox0=\hbox{$\textstyle\rm T$}\hbox{\hbox
to0pt{\kern0.3\wd0\vrule height0.9\ht0\hss}\box0}}
{\setbox0=\hbox{$\scriptstyle\rm T$}\hbox{\hbox
to0pt{\kern0.3\wd0\vrule height0.9\ht0\hss}\box0}}
{\setbox0=\hbox{$\scriptscriptstyle\rm T$}\hbox{\hbox
to0pt{\kern0.3\wd0\vrule height0.9\ht0\hss}\box0}}}}
\def\bbbs{{\mathchoice
{\setbox0=\hbox{$\displaystyle     \rm S$}\hbox{\raise0.5\ht0\hbox
to0pt{\kern0.35\wd0\vrule height0.45\ht0\hss}\hbox
to0pt{\kern0.55\wd0\vrule height0.5\ht0\hss}\box0}}
{\setbox0=\hbox{$\textstyle        \rm S$}\hbox{\raise0.5\ht0\hbox
to0pt{\kern0.35\wd0\vrule height0.45\ht0\hss}\hbox
to0pt{\kern0.55\wd0\vrule height0.5\ht0\hss}\box0}}
{\setbox0=\hbox{$\scriptstyle      \rm S$}\hbox{\raise0.5\ht0\hbox
to0pt{\kern0.35\wd0\vrule height0.45\ht0\hss}\raise0.05\ht0\hbox
to0pt{\kern0.5\wd0\vrule height0.45\ht0\hss}\box0}}
{\setbox0=\hbox{$\scriptscriptstyle\rm S$}\hbox{\raise0.5\ht0\hbox
to0pt{\kern0.4\wd0\vrule height0.45\ht0\hss}\raise0.05\ht0\hbox
to0pt{\kern0.55\wd0\vrule height0.45\ht0\hss}\box0}}}}
\def\bbbz{{\mathchoice {\hbox{$\sf\textstyle Z\kern-0.4em Z$}}
{\hbox{$\sf\textstyle Z\kern-0.4em Z$}}
{\hbox{$\sf\scriptstyle Z\kern-0.3em Z$}}
{\hbox{$\sf\scriptscriptstyle Z\kern-0.2em Z$}}}}
\def\diameter{{\ifmmode\mathchoice
{\ooalign{\hfil\hbox{$\displaystyle/$}\hfil\crcr
{\hbox{$\displaystyle\mathchar"20D$}}}}
{\ooalign{\hfil\hbox{$\textstyle/$}\hfil\crcr
{\hbox{$\textstyle\mathchar"20D$}}}}
{\ooalign{\hfil\hbox{$\scriptstyle/$}\hfil\crcr
{\hbox{$\scriptstyle\mathchar"20D$}}}}
{\ooalign{\hfil\hbox{$\scriptscriptstyle/$}\hfil\crcr
{\hbox{$\scriptscriptstyle\mathchar"20D$}}}}
\else{\ooalign{\hfil/\hfil\crcr\mathhexbox20D}}%
\fi}}
\def\sq{\ifmmode\squareforqed\else{\unskip\nobreak\hfil
\penalty50\hskip1em\null\nobreak\hfil\squareforqed
\parfillskip=0pt\finalhyphendemerits=0\endgraf}\fi}
\def\squareforqed{\hbox{\rlap{$\sqcap$}$\sqcup$}}
\input{isolatin.sty}
\newcommand{\mci}[1]{\multicolumn{1}{c}{#1}}
\newcommand{\mcii}[1]{\multicolumn{2}{c}{#1}}

\newcommand{\Msol}{\mbox{M$_\odot$}}
\newcommand{\spie}{\mbox{$\sigma_{\rm ext}$}}
\newcommand{\spii}{\mbox{$\sigma_{\rm int}$}}
\newcommand{\spiz}{\mbox{$\sigma_0$}}

\newcommand{\VO}[1]{Villafranca~O-{#1}}
\newcommand{\VB}[1]{Villafranca~B-{#1}}
\newcommand{\alphatk}{\mbox{$\alpha_{\rm J2000}$}}
\newcommand{\deltatk}{\mbox{$\delta_{\rm J2000}$}}
\begin{document}
\pagenumbering{arabic}
\pagestyle{myheadings}
\thispagestyle{empty}
{\flushleft\includegraphics[width=\textwidth,bb=58 650 590 680]{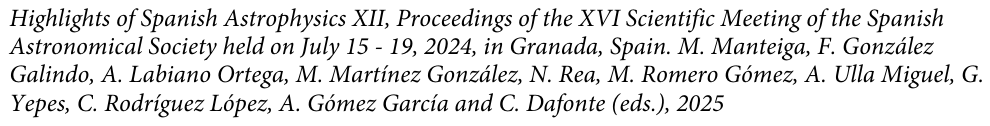}}
\vspace*{-1.0cm}


\begin{flushleft}
{\bf {\LARGE
%
The Villafranca project: Combining Gaia and ground-based surveys to study Galactic OB groups
%
}\\
\vspace*{1cm}
%
J. Maíz Apellániz$^1$,
R. H. Barbá$^2$,
J. A. Molina Lera$^3$,
A. Lambarri Martínez$^{1,4}$,
and 
R. Fernández Aranda$^{1,4,5}$
%
}\\
\vspace*{0.5cm}
%
$^1$ 
Centro de Astrobiolog{\'\i}a, CSIC-INTA, Spain\\
$^2$
Universidad de La Serena, Chile\\
$^3$
Universidad Nacional de La Plata, Argentina\\
$^4$
Universidad Complutense de Madrid, Spain\\
$^4$
University of Crete, Greece\\
%
\end{flushleft}
%
\markboth{
The Villafranca project
}{ 
%
Ma{\'\i}z Apell\'aniz et al.
%
}
\thispagestyle{empty}
\vspace*{0.4cm}
\begin{minipage}[l]{0.09\textwidth}
\ 
\end{minipage}
\begin{minipage}[r]{0.9\textwidth}
\vspace{1cm}
\section*{Abstract}{\small
%
The Villafranca project is studying Galactic stellar groups with OB stars combining information from \textit{Gaia} and ground-based surveys. We summarize the
status of the project and we present its most important results. The Villafranca project has been used to produce a new astrometric calibration for
\textit{Gaia}~(E)DR3, which improves the previous one significantly for bright stars. We have discovered that dynamical interactions among massive stars at a very
young age ($\sim$1~Ma or less) can play a significant interaction in the dynamical evolution of clusters. As a consequence, our current view of the massive-star
IMF may be distorted and the number of free-floating neutron stars and black holes higher than previously considered.
%
\normalsize}
\end{minipage}
%
%


\section{Motivation and summary}

$\,\!$\indent The Villafranca project aims to characterize Galactic stellar groups with OB stars, that is, bound (clusters) or unbound (associations or parts 
thereof) ensembles of young massive stars born together. It is a long-term project designed to study the massive stellar population of the solar neighborhood and
derive its properties: spatial distribution, kinematics, IMF, and relationship with the ISM. Those goals are being achieved by combining astrometric, 
photometric, and spectroscopic information from \textit{Gaia} and ground-based surveys, both spectroscopic such as GOSSS \citep{Maizetal11}, 
LiLiMaRlin \citep{Maizetal19a}, ALS \citep{Pantetal21}, and WEAVE \citep{Jinetal23}, and photometric such as GALANTE \citep{Maizetal21d} and 
MUDEHaR \citep{Holgetal25}.

In this poster we summarize the work in the Villafranca papers, provide the basic information for the 42 groups studied so far in
Table~\ref{maintable}, describe the most significant results, and anticipate our future work.


\section{Papers}

$\,\!$\indent Here we briefly describe the previous and future Villafranca papers as a way to summarize the work done in the project.


\subsection{The proto-Villafranca paper: Method}

$\,\!$\indent In \citet{Maiz19} we established the method used to select the membership and measure the distances to Galactic OB groups. Two objects were studied,
(the later named) \VO{015}~and~O-016.


\subsection{Villafranca I: 16 groups with Gaia DR2}

$\,\!$\indent In Paper I \citep{Maizetal20b} we used \textit{Gaia}~DR2 data to determine the membership and distances to 16 groups with O stars, concentrating in
those with O2-O3.5 objects. We also tied up the identification of the group with the presence of massive stars with accurate spectral types, defined the
nomenclature (\VO{XXX} for OB groups with O stars, \VB{XXX} for those without them), and compared the results with literature values.


\subsection{Villafranca II: 26 groups with Gaia EDR3}

$\,\!$\indent In Paper II \citep{Maizetal22a} we extended the sample to 26 groups with O stars, significantly improved the precision and accuracy by using
\textit{Gaia}~EDR3 data and the astrometric calibration we derived for it, and derived PMS ages for four of the clusters.


\subsection{Villafranca III: The Carina OB1 association}

$\,\!$\indent In Paper III \citep{Molietal24} we study the groups of the Carina~OB1 association. In Table~\ref{maintable} we present our first results.


\subsection{Villafranca IV: The Cepheus spur}

$\,\!$\indent The fourth paper in the series will be on the OB groups of the Cepheus spur (see \citealt{Pantetal21}). We expect to submit it in 2025.


\subsection{Gaia astrometric calibration}

$\,\!$\indent In Villafranca~I we detected a discrepancy between the parallaxes of bright and faint stars in \textit{Gaia}~DR2. With
\textit{Gaia}~(E)DR3 we were able to produce an alternative astrometric calibration \citep{Maizetal21c,Maiz22} that enhances the results of the
Villafranca project and of other analyses that employ \textit{Gaia}~(E)DR3 parallaxes (see subsection~\ref{astrocal}). 


\subsection{Individual cluster papers}            

$\,\!$\indent Once the technique was established with \textit{Gaia}~(E)DR3 data in Villafranca~II, it became possible to use it for individual clusters of 
interest. As a result, the following papers have used it and their targets have been incorporated into the Villafranca project (Table~\ref{maintable}).

\begin{itemize}
 \item \citet{Maizetal22b}: Bermuda cluster, \VO{014}.
 \item \citet{Neguetal22}: Westerlund~1, \VO{033}.
 \item \citet{Ansietal23}: GLS~12\,502 cluster, \VO{034}.
 \item \citet{Putketal23}: NGC~6604, \VO{035}.
 \item \citet{Maizetal24}: Stock~18, \VO{036}.
 \item \citet{MaizNegu25}: Barbá 2, \VB{006}.
\end{itemize}


\begin{figure}
 \centerline{\includegraphics[width=0.70\textwidth]{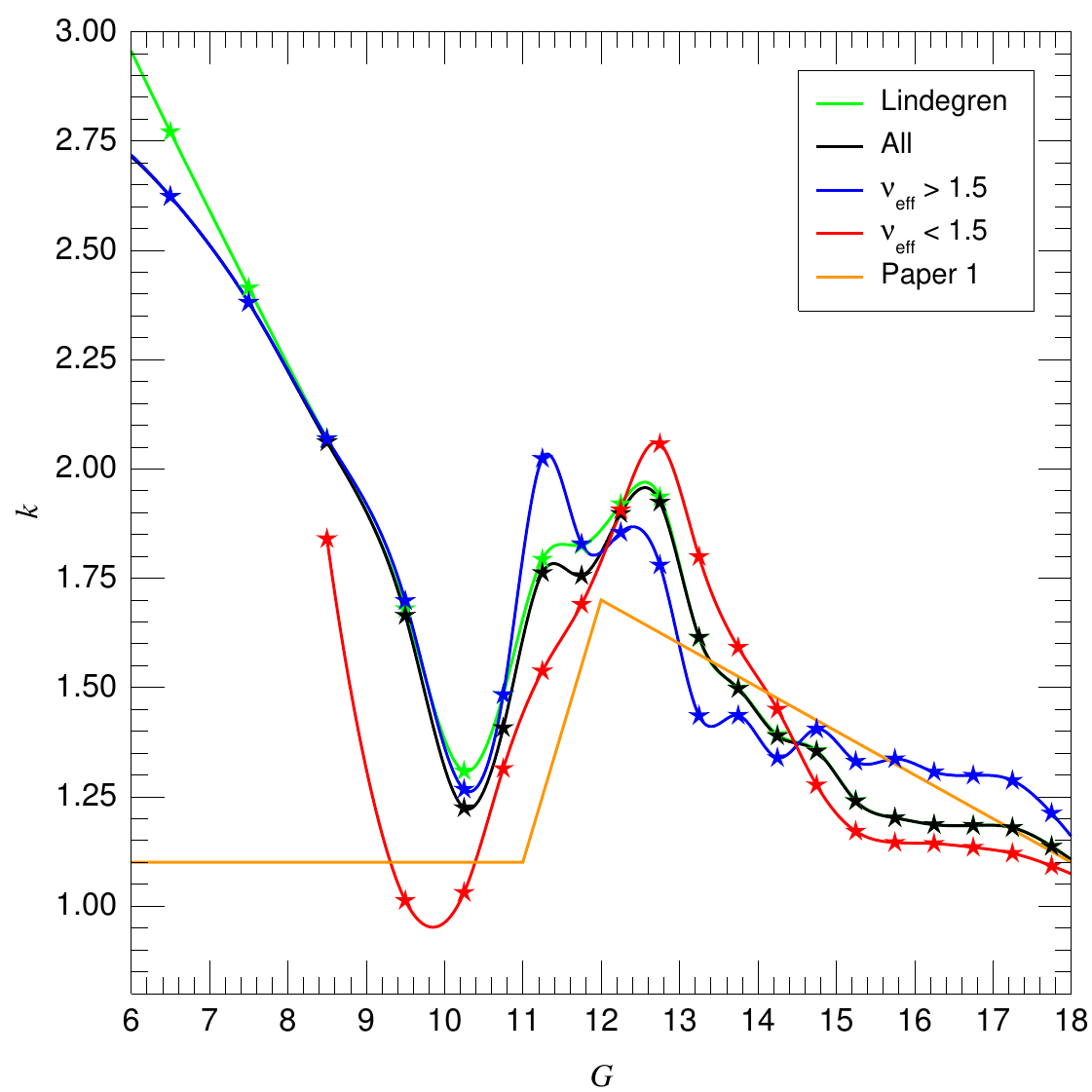}}
 \caption{$k$ as a function of $G$ magnitude for \textit{Gaia}~(E)DR3 using different methods (Eqn.~\ref{spi} and \citealt{Maiz22}).}
 \label{k}
\end{figure}


\begin{table}
\caption{Current Villafranca list of OB groups including the ones in Villafranca~III. Groups in the Carina OB1 association are listed in red.}
\begin{tabular}{llrrr@{}lcl}
 \\
\hline
\mci{Vill.} & \mci{Name}                      & \mci{\alphatk} & \mci{\deltatk} & \mcii{$d$}            & PMS age     & Ref.  \\
\mci{ID}    &                                 & \mci{(deg)}    & \mci{(deg)}    & \mcii{(pc)}           & (Ma)        &      \\
\hline
O-001       & NGC 3603                        & 168.79         & $-$61.26       & 7130&$^{+590}_{-500}$ & ---         & V I   \\
O-002       & \textcolor{red}{Trumpler 14}    & 160.95         & $-$59.56       & 2363&$^{+61}_{-58}$   & ---         & V I   \\
O-003       & \textcolor{red}{Trumpler 16 W}  & 161.09         & $-$59.73       & 2305&$^{+64}_{-61}$   & ---         & V I   \\
O-004       & Westerlund 2                    & 155.99         & $-$57.76       & 4440&$^{+230}_{-210}$ & ---         & V I   \\
O-005       & Pismis 24                       & 261.18         & $-$34.21       & 1642&$^{+30}_{-29}$   & ---         & V I   \\
O-006       & Gum 35                          & 164.68         & $-$61.18       & 7260&$^{+610}_{-520}$ & ---         & V I   \\
O-007       & Cyg OB2-22 cluster              & 308.30         & $+$41.22       & 1620&$^{+30}_{-29}$   & ---         & V I   \\
O-008       & Cyg OB2-8 cluster               & 308.32         & $+$41.31       & 1608&$^{+30}_{-29}$   & ---         & V I   \\
O-009       & M17                             & 275.12         & $-$16.18       & 1696&$^{+41}_{-39}$   & ---         & V I   \\
O-010       & NGC 6193                        & 250.30         & $-$48.76       & 1148&$^{+16}_{-15}$   & ---         & V I   \\
O-011       & Berkeley 90                     & 308.83         & $+$46.84       & 2741&$^{+86}_{-81}$   & ---         & V I   \\
O-012       & NGC 2467                        & 118.18         & $-$26.33       & 4241&$^{+200}_{-180}$ & ---         & V I   \\
O-013       & Sh 2-158                        & 348.43         & $+$61.50       & 2710&$^{+110}_{-100}$ & ---         & V I   \\
O-014       & Bermuda cluster                 & 313.10         & $+$44.40       & 798&$^{+6}_{-6}$      & ---         & V I   \\ 
O-015       & Collinder 419                   & 304.60         & $+$40.78       & 1001&$^{+11}_{-11}$   & ---         & V 0   \\
O-016       & NGC 2264                        & 100.25         &  $+$9.75       & 703&$^{+5}_{-5}$      & 4.0$\pm$2.0 & V 0   \\
O-017       & Heart nebula                    &  38.17         & $+$61.46       & 2075&$^{+44}_{-42}$   & ---         & V II  \\ 
O-018       & Lagoon nebula                   & 271.10         & $-$24.40       & 1234&$^{+16}_{-16}$   & ---         & V II  \\ 
O-019       & Eagle nebula                    & 274.68         & $-$13.79       & 1697&$^{+31}_{-30}$   & ---         & V II  \\ 
O-020       & Rosette nebula                  &  97.98         &  $+$4.94       & 1421&$^{+21}_{-20}$   & ---         & V II  \\
O-021       & NGC 2362                        & 109.68         & $-$24.95       & 1227&$^{+17}_{-16}$   & 5.0$\pm$0.5 & V II  \\
O-022       & NGC 6231                        & 253.54         & $-$41.83       & 1551&$^{+25}_{-24}$   & ---         & V II  \\
O-023       & Orion nebula                    &  83.82         &  $-$5.39       & 390&$^{+2}_{-2}$      & ---         & V II  \\ 
O-024       & $\gamma$ Vel cluster            & 122.38         & $-$47.33       & 336&$^{+1}_{-1}$      & 8.0$\pm$2.0 & V II  \\
O-025       & \textcolor{red}{Trumpler 16 E}  & 161.30         & $-$59.70       & 2311&$^{+58}_{-56}$   & ---         & V II  \\
O-026       & $\sigma$ Ori cluster            &  84.69         &  $-$2.60       & 397&$^{+2}_{-2}$      & 2.0$\pm$0.5 & V II  \\
O-027       & \textcolor{red}{Trumpler 15}    & 161.14         & $-$59.36       & 2354&$^{+61}_{-58}$   & ---         & V III \\
O-028       & \textcolor{red}{Collinder 228}  & 160.88         & $-$59.97       & 2339&$^{+57}_{-54}$   & ---         & V III \\
O-029       & \textcolor{red}{Collinder 232}  & 161.24         & $-$59.53       & 2322&$^{+59}_{-56}$   & ---         & V III \\
O-030       & \textcolor{red}{Bochum 11}      & 161.80         & $-$59.96       & 2327&$^{+56}_{-54}$   & ---         & V III \\
O-031       & \textcolor{red}{NGC 3324}       & 159.45         & $-$58.64       & 2417&$^{+63}_{-60}$   & ---         & V III \\
O-032       & \textcolor{red}{Loden 153}      & 158.47         & $-$58.14       & 2421&$^{+65}_{-62}$   & ---         & V III \\
O-033       & Westerlund 1                    & 251.76         & $-$45.85       & 4240&$^{+260}_{-230}$ & ---         & N22   \\
O-034       & GLS \num{12502} cluster         & 338.69         & $+$58.30       & 3890&$^{+190}_{-170}$ & ---         & A23   \\
O-035       & NGC 6604                        & 274.53         & $-$12.23       & 1941&$^{+38}_{-36}$   & ---         & P23   \\
O-036       & Stock 18                        &   0.40         & $+$64.63       & 2910&$^{+100}_{-100}$ & $\sim$1.0   & M24   \\
\hline
B-001       & \textcolor{red}{NGC 3293}       & 158.95         & $-$58.24       & 2335&$^{+58}_{-56}$   & ---         & V III \\
B-002       & \textcolor{red}{Bochum 10}      & 160.55         & $-$59.13       & 2378&$^{+61}_{-58}$   & ---         & V III \\
B-003       & \textcolor{red}{ASCC 62}        & 162.74         & $-$59.94       & 2403&$^{+59}_{-57}$   & ---         & V III \\
B-004       & \textcolor{red}{IC 2581}        & 156.85         & $-$57.63       & 2463&$^{+65}_{-62}$   & ---         & V III \\
B-005       & \textcolor{red}{Ruprecht 90}    & 157.95         & $-$58.24       & 2544&$^{+71}_{-67}$   & ---         & V III \\
B-006       & Barbá 2                         & 166.09         & $-$61.75       & 7390&$^{+650}_{-550}$ & ---         & M25   \\
\hline
\end{tabular}
\label{maintable}                  
\end{table}


\section{Results}

$\,\!$\indent Here we present the most important results of the Villafranca project.


\subsection{Group definitions and distances}

$\,\!$\indent The main result of the project is the definition of the stellar groups and the determination of their distances. At this point we have 
analyzed a total of 42~OB groups, of which 36 include O-type stars (Table~\ref{maintable}) and the number will keep growing in the future, starting with Paper~IV.
As discussed in Paper~I, the \textit{Gaia} distances provide a significant improvement over previous methods.


\subsection{Measuring distances with Gaia DR3 properly}\label{astrocal}

$\,\!$\indent A number of recent papers have published \textit{Gaia}-based distances for stellar groups but one has to be careful to avoid systematic biases
(accuracy) and to correctly estimate random uncertainties (precision). For that reason, the Villafranca project has established an independent and improved
astrometric calibration of \textit{Gaia} data, of which the most relevant points are:

\begin{itemize}
 \item \textbf{The catalog (internal, \spii) \textit{Gaia}~(E)DR3 parallax uncertainties are significantly underestimated with respect to the 
       real (external, \spie) parallax uncertainties}. Using:
       \begin{equation}
         \spie^2 = k^2\spii^2 + \spiz^2
         \label{spi}
       \end{equation}
       \citep{Lindetal18b}, $k$ is larger than 1.0 and as high as 3.0 for bright stars (\citealt{Fabretal21a,Maiz22}, Fig.~\ref{k}).
 \item \textbf{\textit{Gaia} parallaxes have a substantial angular correlation}, hence the \spiz\ term in Eqn.~\ref{spi}
       \citep{Lindetal18a}. For \textit{Gaia}~(E)DR3, \spiz\ is 10.3~$\mu$as \citep{Maizetal21c}.
 \item \textbf{The existence of an angular correlation imposes a limit on the distance uncertainty for clusters achievable with \textit{Gaia}}. For 
       \textit{Gaia}~(E)DR3, this amounts to the distance in kpc as a percent e.g. 1\% at 1~kpc and 3\% at 3~kpc. If you see a published distance with a better 
       uncertainty, check if the authors are considering the angular correlation or not.
 \item \textbf{\textit{Gaia}~(E)DR3 parallaxes have a significant but correctable parallax bias that depends on magnitude, color, and position}
       \citep{Lindetal21b}. For bright stars $G < 13$, the \citet{Maiz22} parallax bias works better.
 \item \textbf{Distances derived from parallaxes depend on the prior you use} \citep{Maiz01a,Lurietal18}. Make sure you are using one appropriate for your 
       sample. 
\end{itemize}


\begin{figure}
 \centerline{\includegraphics[width=0.55\textwidth,valign=t]{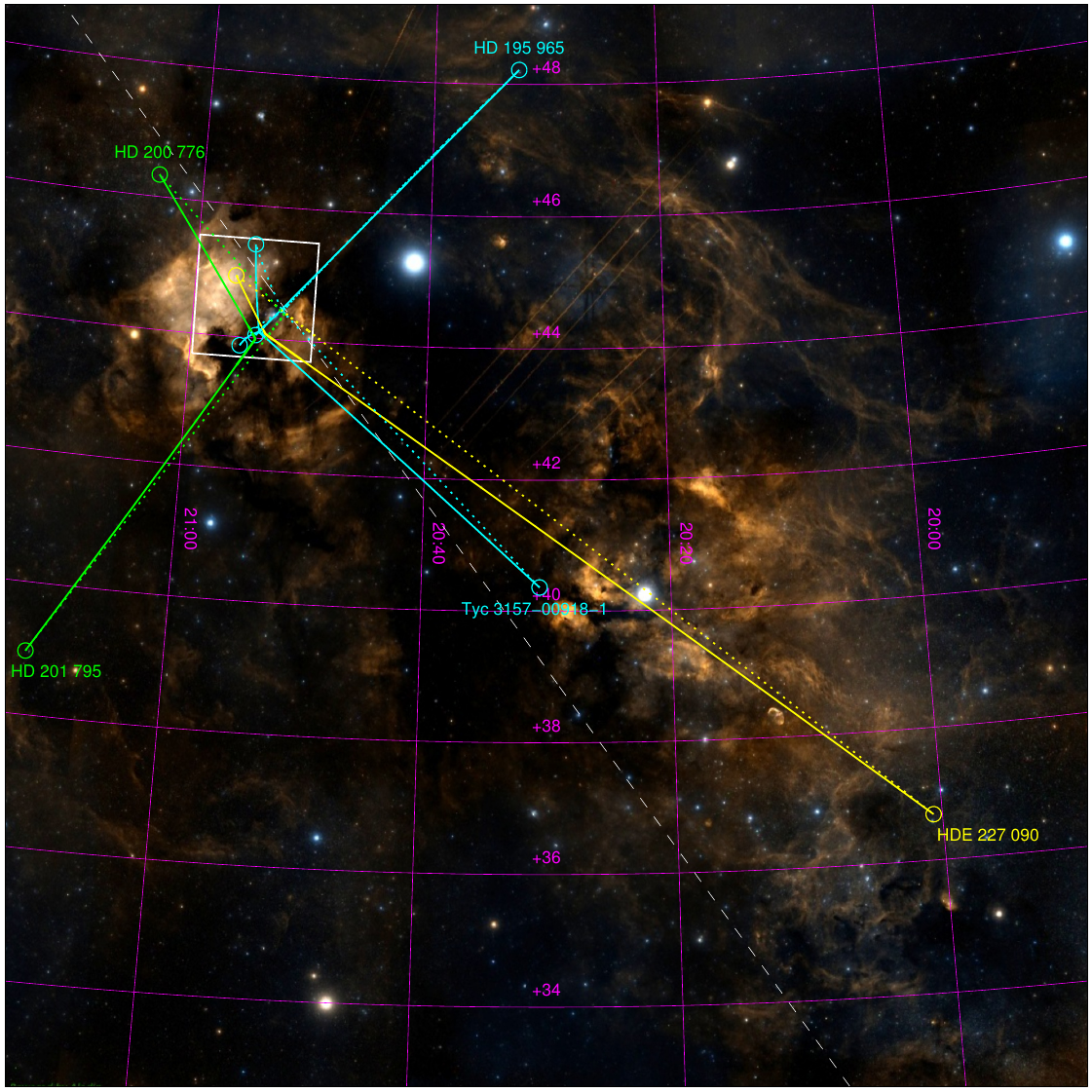} \
             \includegraphics[width=0.55\textwidth,valign=t]{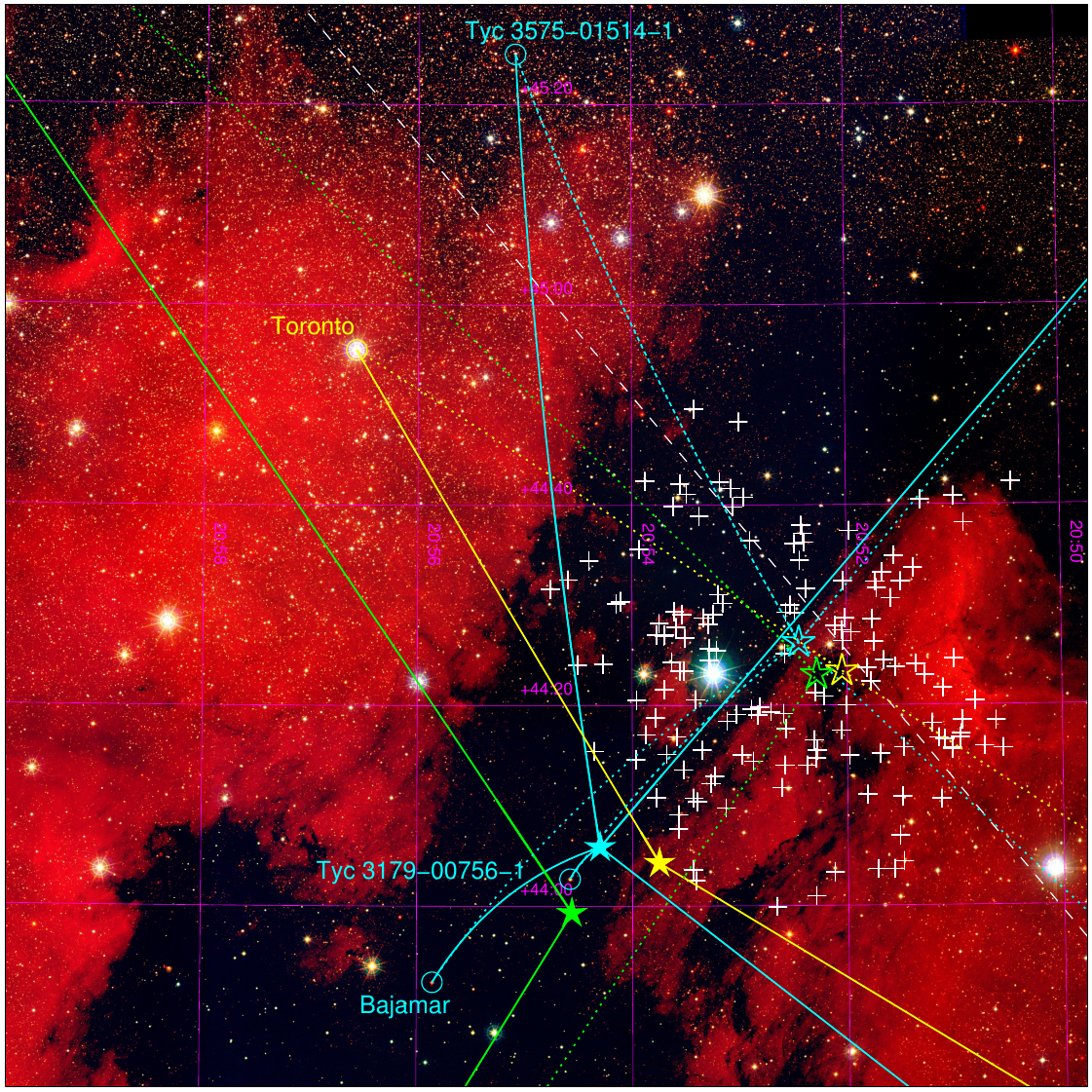}}
 \caption{(left) DSS2 image with the three ejection events in the Bermuda cluster, color-coded in cyan (Bajamar), yellow (Toronto), and green (HD~201\,795).
         Colored solid lines show a representative trajectory for each system in the Sun's LSR and colored short-dashed lines the equivalent after subtracting the motion of 
         the Bermuda cluster. The long-dashed line shows the Galactic equator. (right) Equivalent GALANTE three-color mosaic of the square region in the left panel. Stars 
         mark the location of each event in the Sun's LSR (filled) and after subtracting the motion of the Bermuda cluster (non-filled).}
 \label{Bermuda_events}
\end{figure}


\subsection{Expanding clusters and ejected stars}

$\,\!$\indent We have detected a significant dynamical evolution in their first 1-2~Ma for four Villafranca clusters:

\begin{itemize}
 \item The Bermuda cluster in the North America nebula experienced three stellar dynamical interaction events 1.9,~1.6,~and~1.5~Ma ago that can be traced by
       the runaway/walkaway stars they produced (\citealt{Maizetal22b}, Fig.~\ref{Bermuda_events}). Most of the massive stars have been ejected, leaving behind an
       unbound orphan cluster with an expansion time scale compatible with the dynamical interaction events.
 \item NGC~2467 has also been orphaned by an ejection event that took place 0.4~Ma ago and that left the cluster without its two most massive systems, 
       HD~64\,568 and HD~64\,315~A,B (Villafranca~II).
 \item Stock~18 has retained its only O star but ejected most of its massive stars in one or several events that took place $\sim$1.0~Ma ago
       \citep{Maizetal24}.
 \item Trumpler~16, the cluster that hosts $\eta$~Car, is divided into two subclusters (\VO{003}~and~O-025) that have been moving away from each other for the
       last 1.0-1.5~Ma, with signs of additional internal expansion \citep{Molietal24}.
\end{itemize}

In addition, at the location of the Orion nebula cluster several massive stars were ejected 2.5~Ma ago, likely dissolving its proto-cluster \citep{Hoogetal00}.
These results indicate that \textbf{dynamical interactions among massive stars at a very young age (1~Ma or less) can play a significant role in the dynamical
evolution of clusters} \citep{OhKrou16}. Such interactions take place before the classical gas loss mechanism of \citet{LadaLada03}.

\subsection{The IMF and the PDMF}

$\,\!$\indent For at least three of the clusters above (Bermuda, NGC~2467, and Stock~18), the PDMF at an age of 3~Ma (before any SN explosions) will look 
quite different to the IMF: The PDMF is close to canonical \citep{Krou01} while the IMF is top-heavy. These are low-mass clusters (several hundred \Msol) that
should not form several (if any) O stars yet they still have (the most massive O star within 1~kpc was actually formed in the Bermuda cluster). Therefore,
\textbf{the effect of early dynamical interactions can yield a distorted view of the IMF and the current Galactic massive-star formation rate may be higher than
previously thought.}

\hspace{5mm} As the (previously unaccounted for) additional massive stars are runaways/walkaways, \textbf{the number of free-floating neutron stars and black 
holes may be significantly higher than previously considered.}


\section{Future work}

\begin{itemize}
 \item We will keep adding groups to the catalog.
 \item We will reprocess the existing groups once new \textit{Gaia} data releases become available.
 \item Data from ground-based surveys such as GALANTE and WEAVE will be incorporated.
 \item The results will be used to study Galactic structure, the origin of runaway/walkaway stars, mechanisms of cluster formation, and the IMF.
\end{itemize}


\bibliographystyle{aa}
\bibliography{general}

\end{document}